%% file: main.tex
\documentclass[11pt]{article}

\usepackage[letterpaper,margin=1in]{geometry}

\usepackage{amsmath}
\usepackage{amsfonts}
\usepackage{amssymb}
\usepackage{amsthm}

\usepackage[square,numbers,sort,compress]{natbib}
\usepackage{authblk}

\usepackage{graphicx} % Required for inserting images
\usepackage{enumerate}
\usepackage[linesnumbered,commentsnumbered,ruled]{algorithm2e}
\usepackage{booktabs}
\usepackage{multirow}
\usepackage{multicol}
\usepackage{threeparttable}

\usepackage{url}
\usepackage{hyperref}
\usepackage[hyphenbreaks]{breakurl}
\usepackage{cleveref}

\newcommand{\cbr}[1]{\left\{ #1 \right\}}

\providecommand{\keywords}[1]{{\noindent\small\textbf{\textit{Keywords---}} #1}}

\title{Static Batching of Irregular Workloads on GPUs: Framework and Application to Efficient MoE Model Inference}

\author{Yinghan Li\thanks{Equal contribution.} }
\author{Yifei Li$^\ast$}
\author{Jiejing Zhang}
\author{Bujiao Chen}
\author{Xiaotong Chen}
\author{Lian Duan}
\author{Yejun Jin}
\author{\par Zheng Li}
\author{Xuanyu Liu}
\author{Haoyu Wang}
\author{Wente Wang}
\author{Yajie Wang}
\author{Jiacheng Yang}
\author{\par Peiyang Zhang}
\author{Laiwen Zheng}
\author{Wenyuan Yu}
\affil{Alibaba Group}
\affil{\texttt{\{lyh238099,lyf383659,jiejing.zjj,wenyuan.ywy\}@alibaba-inc.com}}
\date{}

\begin{document}
\maketitle

%% Abstract
\input{tex/abstract}

%% Keywords
\keywords{batching, irregular workload, MoE, LLM inference, GPGPU}

%% Contents
\input{tex/introduction}
\input{tex/background}

\input{tex/framework}
\input{tex/moekernel}

\input{tex/evaluation}

\input{tex/conclusion}

%% Bibliography
\bibliography{bib/ref.bib}

\end{document}

%% file: tex/abstract.tex
\begin{abstract}
It has long been a problem to arrange and execute irregular workloads on massively parallel devices.
We propose a general framework for statically batching irregular workloads into a single kernel with a runtime task mapping mechanism on GPUs.
We further apply this framework to Mixture-of-Experts (MoE) model inference and implement an optimized and efficient CUDA kernel.
Our MoE kernel achieves up to 91\% of the peak Tensor Core throughput on NVIDIA H800 GPU and 95\% on NVIDIA H20 GPU.
\end{abstract}

%% file: tex/introduction.tex
\section{Introduction}
\label{sec:intro}

Resource utilization is one of the key factors in fully exploiting the computing power of massively parallel devices, including GPUs.
As a common method to improve utilization and reduce overhead, the benefit of the batching technique should never be underestimated~\cite{knk_batch_2022,kpp_batch_2021,li_topk_2024}.
In most cases, it is handy to batch regular workloads that share the same type and size, which also have similar amounts of computation and memory access.
For example, in the CUDA programming model, this kind of regular workloads can be conveniently batched along an additional thread block or grid dimension~\cite{cuda_guide}.

However, irregular workloads do not naturally fit into this scheme.
Irregular workloads may show one or more of the following characteristics that prevent regular batching~\cite{chen_atos_2023}: variable amounts of computation, special memory access patterns, control flow divergence, etc.
Moreover, \emph{heterogeneous} workloads almost raise the difficulty of batching to an unreachable level.
Here, by heterogeneous, we refer to workloads of different types of operations, e.g., some of the workloads are reduction, while others are element-wise operations.

Irregular workloads are often managed in a \emph{task-parallel} fashion instead of batching, where an individual workload is regarded as a task, and all tasks are dynamically scheduled~\cite{chen_atos_2023,tzeng_irregular_2010}.
The commonly used CUDA stream interface is an intra-process example of this scheme~\cite{cuda_guide}, while inter-process task-parallel is also available via tools such as CUDA MPS~\cite{cuda_mps, duan_muxserve_2024}.
The task-parallel technique is generally effective for irregular workloads, but it can be suboptimal for specified types of workloads due to coarse-grained scheduling, as well as the overhead of scheduling and launching~\cite{li_tiling_2019}.

Nevertheless, for specific types of workloads, batching rather than task parallelism is still the preferred method, for example, \emph{General Matrix Multiplication} (GEMM), which is one of the most important tasks for massively parallel devices.
Given multiple independent GEMM workloads, it is natural to batch them together for better resource utilization.
\emph{Batched GEMM} computes a batch of GEMMs of the same shape~\cite{nv_cublas_2024}, while \emph{grouped GEMM} is proposed for GEMMs of different input and output shapes~\cite{nv_cublas_2024,nv_groupedgemm_2024,triton_groupedgemm_2024}.
There is also a two-phase GEMM batching framework that pre-computes the mapping between GEMM tiles and thread blocks on the host, and passes the mapping as a parameter to the kernel, reducing the overhead of task scheduling on the device~\cite{li_tiling_2019}.
Despite these methods for batching GEMMs, there is no common framework for statically batching general irregular workloads.

One of the most cutting-edge irregular workload applications is the \emph{Mixture-of-Experts} (MoE) model inference.
Nowadays, the weights of Large Language Models (LLM) have reached enormous sizes.
For example, Llama 3.1 provides a 405B-parameter model~\cite{meta_llama31_2024}.
To train and infer LLMs more effectively, the MoE technique is introduced, which divides the neural network weights into partitions called \emph{experts}.
For each token, only a dynamic subset of experts is activated, reducing the amount of computation and memory access at runtime while preserving the capability of large models~\cite{fedus_reviewsparseexpertmodels_2022,jiang_mixtralexperts_2024}.

The core operation of the MoE layer is the multiplication of the expert weight and the token tensor.
Different tokens are routed to different subsets of experts~\cite{nv_moe_2024}, i.e., different expert weights are multiplied by different token tensors, making the MoE model inference a kind of irregular workload.
The state-of-the-art is to implement MoE as a grouped GEMM~\cite{nv_groupedgemm_2024}.
On the one hand, this inherits the potential defects of grouped GEMM, such as all tasks sharing the same tiling strategy, which can degrade performance if the GEMM shapes vary greatly between tasks.
On the other hand, the grouped GEMM API requires extra data preparation and prevents possible optimizations to reduce data duplication.

In this paper, we propose a general framework for statically batching irregular workloads on GPUs, and apply it to MoE implementation to achieve highly efficient MoE model inference.
Conceptually, a batch consists of multiple \emph{tasks}, and each task can be further partitioned into \emph{tiles}. 
This framework constructs on the host a compressed mapping from the thread block index to the pair of task index and tile index inside the task.
Then on the device, the kernel efficiently decompresses the mapping, and dispatches the corresponding workload of the target tile to each thread block.

To apply this framework to MoE, we recognize a special case where some experts may not receive any token at all in a particular inference step.
Thus, we further propose an optimization for batches containing empty tasks by introducing an extra stage of mapping into our framework.
Additionally, we introduce a token index array for each expert to eliminate duplicate copies of token tensors.
Several general optimizations for GEMM are also used to achieve maximum computing power on the latest GPUs.

In summary, we claim the following contributions.
\begin{itemize}
    \item We propose a general framework for statically batching irregular workloads on GPUs.
    \item We extend this framework to handle batches containing empty tasks and apply it to MoE model inference.
    \item We implement a highly efficient CUDA kernel for MoE inference and achieve up to 95\% of the peak Tensor Core throughput on the latest NVIDIA GPUs.
\end{itemize}

%% file: tex/background.tex
\section{Background and Motivation}
\label{sec:background}

\subsection{Batching GEMMs on GPUs}
\label{sec:background:batching}

It is the regular case where the GEMM takes tensors with more than 2 dimensions as inputs, and the additional outer dimensions can be flattened as a ``batch dimension" while calculating the multiplication of the inner-most 2D matrices.
Libraries provide static batching APIs for this scenario, e.g., the \emph{batched GEMM} APIs of cuBLAS~\cite{nv_cublas_2024}.

There is a more irregular case, called \emph{grouped GEMM}~\cite{nv_cublas_2024,nv_groupedgemm_2024,triton_groupedgemm_2024}, where the GEMM tasks are of different input and output shapes.
The basic idea is that GEMM is computed in tiles on GPUs.
Grouped GEMM launches a constant number of thread blocks, and dynamically schedules unfinished tiles onto idle blocks~\cite{triton_groupedgemm_2024}. 
One defect of grouped GEMM is that the problem description, including the GEMM shapes and the group size, is loaded inside the kernel, and the tile mapping and scheduling are dynamically executed inside the kernel.
These add overheads to the GEMM computation when the total number of tiles is large.
Another major defect is that all tasks share the same tiling strategy in grouped GEMM.
This can degrade performance if the GEMM shapes vary greatly between tasks, because too large tiling results in a waste of computing power, while too small tiling suffers from low computing power utilization due to low computational intensity~\cite{li_tiling_2019}.

A two-phase GEMM batching framework is also proposed to support different tiling strategies within a batch, which pre-computes the mapping between tiles and thread blocks on the host, reducing the kernel overhead on the device as well~\cite{li_tiling_2019}.
However, it introduces as a kernel parameter a mapping array whose length equals the total number of thread blocks.
Thus, if the number of blocks is large, there will be significant overhead in copying the array from host to device.
In addition, when device threads access this array inside the kernel, the cache hit rate is relatively low due to poor data locality.

\subsection{MoE Model Inference}
\label{sec:background:moe}

The key structure of MoE models is the MoE layer, which first selects the subset of experts for a token, then computes the products of the token tensor and each selected expert weight tensor, and finally sums them up as the output.
Generally, multiple tokens are parsed in a batch to improve throughput.
The subsets of experts for different tokens, however, are usually different~\cite{nv_moe_2024}.

There are two common types of parallelism in implementing the MoE layer on GPUs: \emph{tensor parallelism} (TP) and \emph{expert parallelism} (EP)~\cite{deepseek_v3_2024, ms_dsmoe_2022}.
TP splits each expert weight into several parts, and each GPU holds a part of every expert weight.
In terms of EP, a subset of experts reside on each GPU.
For both TP and EP with more than one expert per GPU, the MoE computation is an irregular workload from the perspective of each GPU, since the input token tensor for each expert is distinct not only in data but also in shape.
In practice, TP and EP can be combined to better scale with hardware resources.

As for implementation, a na\"ive way is to use a for loop to compute GEMMs one by one instead of batching.
This method is adopted by systems relying more on EP with only a few experts per GPU, for example, the DeepSpeed-MoE inference~\cite{ms_dsmoe_2022}.
Heavy EP requires more GPUs per expert group, which puts more pressure on resources.
The state-of-the-art implementation is to convert MoE into grouped GEMM~\cite{nv_moe_2024}.
However, as mentioned, grouped GEMM may involve overhead due to dynamic tile scheduling.
Moreover, for MoE inference, a token can be routed to multiple experts, resulting in duplicate copies of token tensors to prepare contiguous input tensors for grouped GEMM APIs.

\subsection{Motivation}
\label{sec:background:motiv}

Although several batching methods have been proposed for GEMM to adapt to different types of scenarios, including both regular and irregular workloads, there is still no common framework for batching general irregular tasks on GPUs.
In addition, even the existing irregular GEMM batching methods suffer from various kinds of overheads and defects as discussed in \Cref{sec:background:batching}.

Meanwhile, cutting-edge irregular workloads desire more efficient batching methods. 
For example, MoE inference will greatly benefit from an implementation that overcomes the defects of the SOTA based on grouped GEMM, especially when the numbers of tokens routed to different experts vary greatly in an inference step, which is a common problem called \emph{unbalanced expert load} for MoE models.
Thus, we propose a static batching framework for general irregular workloads, and apply it to MoE inference, achieving almost full peak Tensor Core throughput on the latest GPUs.

%% file: tex/framework.tex
\section{Irregular Workload Batching Framework}
\label{sec:framework}

One of the key components of massively parallel algorithms is task mapping.
Generally speaking, a task is partitioned into several sub-tasks which are mapped onto parallel execution units on the hardware.
Such sub-tasks are sometimes called tiles, especially for GEMM-like tasks where a tile directly corresponds to a submatrix.
From the perspective of CUDA programming model, a tile is usually mapped onto and computed by a thread block.

In this section, we introduce a mapping mechanism with which any thread inside a thread block can efficiently find out which tile it should handle.
Atop this mapping mechanism, we propose a framework for batching any kind of irregular workload on GPUs.

\subsection{Compressed Task Mapping Mechanism}
\label{sec:framework:mapping}

To describe the mapping between thread blocks and GEMM tiling, the prior art uses an auxiliary array whose length equals the number of thread blocks~\cite{li_tiling_2019}.
This na\"ive idea suffers from significant overhead when a number of thread blocks are launched given large input sizes.
In face of this challenge, we design a compressed task mapping mechanism with reduced auxiliary array size, which not only degrades potential copy overhead, but also improves data locality as well as cache hit rate.

First, the mapping itself should be defined.
Specifically, each task in the batch can be partitioned into multiple tiles, with each tile corresponding to a thread block conceptually.
Given a thread block index, the mapping should tell the target task index and tile index inside this task.
On the contrary, for each task, this mapping can determine how many tiles (thread blocks) it requires. 
Then, a compressed auxiliary array, namely \texttt{TilePrefix}, is constructed, containing the inclusive prefix sum of the number of tiles required by each task, as shown in \Cref{alg:build-prefix}.
To be mentioned, this can be either pre-computed on the host and then copied to the device, or directly generated on the device.
Since the length of \texttt{TilePrefix} equals the number of tasks, which is much smaller than the number of thread blocks in general, the copy overhead is very small.
Besides, in practice, the prefix sum can be computed with parallel implementation.

\begin{algorithm}[ht]
    \caption{Build \texttt{TilePrefix} array\label{alg:build-prefix}}

    \small
    \SetAlgoLined
    \SetKwComment{Comment}{// }{}

    \SetKwInOut{Input}{Input}
    \SetKwInOut{Output}{Output}
    % \SetKwFor{For}{for (}{) $\lbrace$}{$\rbrace$}

    \Input{$N$ tasks $\cbr{T_1, \dots, T_N}$, function $\nu(\cdot)$ returning the number of tiles required by a task}
    \Output{Array \texttt{TilePrefix}}

    Initialize array \texttt{TilePrefix} of size $N$\;
    \For{$i\gets 1$ \KwTo $N$}{
      $\texttt{TilePrefix}[i]\gets \sum_{j=1}^i{\nu(T_j)}$\;
    }
\end{algorithm}

With \texttt{TilePrefix} well prepared on the device, inside the kernel, we propose an algorithm to uncompress the mapping compressed in \texttt{TilePrefix}.
Given the index of the current thread block, the idea is to find the first task whose inclusive prefix sum is no less than this index.
Then this block must belong to that task.
The SIMT algorithm is shown in \Cref{alg:compute-mapping}.

\begin{algorithm}[ht]
    \caption{Compute the task mapping with a warp\label{alg:compute-mapping}}

    \small
    \SetAlgoLined
    \SetKwComment{Comment}{// }{}

    \SetKwInOut{Input}{Input}
    \SetKwInOut{Output}{Output}

    \Input{Array \texttt{TilePrefix}, thread block index $B$}
    \Output{Task index $h$, tile index $l$}

    \Comment{Using a warp}
    $t\gets$ thread index\;
    $p\gets B \ge \texttt{TilePrefix}[t]$\Comment*[l]{$p$ is a boolean value}
    $mask\gets$ warp vote of $p$\;
    $h\gets$ population count of $mask$\;
    $k\gets0$\;
    \If{$h > 0$} {
        $k\gets\texttt{TilePrefix}[h-1]$\;
    }
    $l\gets B - k$\;
\end{algorithm}

Here, we provide some details about \Cref{alg:compute-mapping}.
This SIMT algorithm is executed by a warp containing a bunch of consecutive threads, typically 32 threads, which is the minimum scheduling unit of GPUs.
Warp voting is a mechanism that generates an integer mask whose $i$-th bit is set if the boolean value held by thread $i$ is true.
Population count is a mechanism provided by most architectures that computes the number of set bits in an integer.
If using only one warp for computation, after the warp finishes finding the task index and tile index, these values can be broadcast to other warps via shared memory.
Also, it is possible to let all warps execute the algorithm above, which may even show better performance, because the size of \texttt{TilePrefix} is small and the L1 cache hit rate will be high.

If $N$ is smaller than the warp size, \texttt{TilePrefix} will need to be padded up to the warp size by repeating its last element or padding with the maximum possible value.
Note that \Cref{alg:compute-mapping} works for $N$ no more than the warp size.
For larger $N$, we can simply let each warp loop this algorithm several times to scan the whole \texttt{TilePrefix} array.
And for even larger $N$, e.g., $N=512$, we can build 2-level or multi-level \texttt{TilePrefix} arrays, which is omitted in this paper.

% ------------------------------

\subsection{Batching Irregular Tasks}
\label{sec:framework:batching}

Different tasks inside a batch can have different tiling strategies.
Further, the tasks themselves can be heterogeneous, which means that two tasks can be different types of operations, e.g., one is GEMM and the other is reduction sum.
However, as long as the tile partition scheme can be determined before kernel launch, we are able to statically batch any kind of irregular workload by implementing \Cref{alg:compute-mapping} as a device funciton, namely \texttt{mapping}.

Usually, different types of tasks are implemented as separate kernels, i.e., global functions.
In the classic task-parallel scheme, tasks are dynamically scheduled, and their kernels are launched independently on their own streams~\cite{tzeng_irregular_2010}.
While for static batching, the original global functions need to be rewritten as device functions so that they can be integrated together into a single kernel.
Suppose the number of different types of tasks (or different tiling strategies) is $K$, and the corresponding device functions are $\texttt{taskFunc}_1,\dots,\texttt{taskFunc}_K$.
The batching framework is shown in \Cref{alg:framework} and can be implemented as a global function.

\begin{algorithm}[ht]
    \caption{Batching framework\label{alg:framework}}

    \small
    \SetAlgoLined
    \SetKwComment{Comment}{// }{}

    \SetKwInOut{Input}{Input}
    \SetKwInOut{Output}{Output}

    \SetKw{Break}{break}

    \Input{$N$ tasks $\cbr{T_1, \dots, T_N}$, task parameters $\cbr{p_1, \dots, p_N}$, array \texttt{TilePrefix}, thread block index $B$}
    % \Output{}

    $h,\ l \gets \texttt{mapping}(\texttt{TilePrefix},\ B)$\;
    \For{$i\gets 1$ \KwTo $K$}{
        \If{task type of $\ T_h$ is $i$} {
            $\texttt{taskFunc}_i(l,\ p_h)$\;
            \Break\;
        }
    }
\end{algorithm}

%% file: tex/moekernel.tex
\section{Application to MoE Model Inference}
\label{sec:moe}

As mentioned, MoE model inference can be converted into batching irregular GEMMs by regarding the matrix multiplication of each expert as a task.
These GEMMs differ from each other in that their input and output shapes can be quite different due to the unbalanced expert load.
To better utilize the computing power, these GEMMs can be categorized into several pre-defined tiling strategies.
Generally speaking, GEMMs with large input and output sizes prefer large tiles to improve computational intensity.
Each individual tiling strategy can be implemented as a device function. 
Thus, the MoE inference is expected to fit into the framework in \Cref{alg:framework}.

However, due to the unbalanced expert load, it is possible that no token is routed to an expert in a particular inference step, resulting in an empty task.
The mapping algorithm in \Cref{alg:compute-mapping} cannot handle potential empty tasks, so we propose a two-stage mapping to extend the aforementioned framework.
Atop the extended framework, we implement a high-performance MoE inference kernel that statically batches the MoE expert GEMMs, each with the best tiling strategy.
We also introduce token index arrays to eliminate the copy overhead of token tensors.
In addition, we leverage several common GEMM optimizations in our kernel implementation to maximize the performance achieved on the latest GPUs.

\subsection{Extended Batching Framework for Potential Empty Tasks}
\label{sec:moe:extension}

The mapping ``thread block index $\mapsto$ task index" in \Cref{alg:compute-mapping} fails when some tasks in the batch are empty, i.e., the number of tiles they require is zero.
To extend the batching framework to handle potential empty tasks in MoE inference, we add an extra stage of mapping: ``thread block index $\mapsto$ non-empty task index $\mapsto$ real task index".

Similar to the setting in \Cref{sec:framework}, suppose there are $N$ tasks $\tau=\cbr{T_1,\dots,T_N}$, among which $M\le N$ tasks are non-empty, denoted $\eta=\cbr{S_1,\dots,S_M}\subseteq\tau$.
There exists an injection $\sigma:[M]\to[N]$, such that $\forall i\in[M], S_i = T_{\sigma(i)}$.
Here, $\sigma$ is the mapping from a non-empty task index to the real task index.
We only build the \texttt{TilePrefix} array for non-empty tasks, and then extend the framework in \Cref{alg:framework} into \Cref{alg:framework-ext}.

\begin{algorithm}[ht]
    \caption{Extended batching framework\label{alg:framework-ext}}

    \small
    \SetAlgoLined
    \SetKwComment{Comment}{// }{}

    \SetKwInOut{Input}{Input}
    \SetKwInOut{Output}{Output}

    \SetKw{Break}{break}

    \Input{$N$ tasks $\cbr{T_1, \dots, T_N}$ among which $M$ tasks $\cbr{S_1,\dots,S_M}$ are non-empty, task parameters $\cbr{p_1, \dots, p_N}$, array \texttt{TilePrefix} for non-emtpy tasks, thread block index $B$}
    % \Output{}

    $h,\ l \gets \texttt{mapping}(\texttt{TilePrefix},\ B)$\Comment*[l]{non-empty task mapping}
    $\Tilde{h}\gets\sigma(h)$\Comment*[l]{real task mapping}
    \For{$i\gets 1$ \KwTo $K$}{
        \If{task type of $\ T_{\Tilde{h}}$ is $i$} {
            $\texttt{taskFunc}_i(l,\ p_{\Tilde{h}})$\;
            \Break\;
        }
    }
\end{algorithm}

To apply this extended framework to the MoE model inference, simply let each expert be a task.
The task parameters $\cbr{p_1, \dots, p_N}$ contain the weight and other necessary information of the experts.
In each inference step, after the token route, we can decide which experts are non-empty in this step and construct the mapping $\sigma$.
The \texttt{TilePrefix} is built for the non-empty experts, and the framework in \Cref{alg:framework-ext} can be applied.

\subsection{Expert Ordering}
\label{sec:moe:order}

In addition to the empty expert problem, the load balance of experts also influences the performance of the MoE kernel.
The impact on performance comes from two sides: the tiling strategy, and the resource utilization.
The challenge of the tiling strategy can be effectively solved with our proposed batching framework.
However, resource utilization is more of an inherited challenge.

Generally speaking, an expert with a large number of tokens, namely busy experts, corresponds to a compute-bound task, while an expert with only a few tokens, namely non-busy experts, corresponds to a memory-bound task.
If a wave of thread blocks are all assigned compute-bound tasks, the memory bandwidth may be somehow wasted, and vice versa.
However, there are still opportunities for optimization.
We notice that during inference, the MoE workloads are sometimes not so large that the entire wave is occupied by thread blocks of only one task.
That is to say, it is possible to mix different tasks in a wave to balance the use of computing power and memory bandwidth.

The basic idea is to interleave busy experts with non-busy experts so that a wave of thread blocks optimally contains both compute-bound and memory-bound tasks.
The key is to design an expert ordering algorithm given the amount of expert workloads.
We try some simple strategies, including alternating busy and non-busy experts, and arranging busy experts in a half-interval manner.
In practice, the half-interval strategy shows better performance.
However, the best ordering of experts is an NP problem.
We leave any further discussion on this topic for future work.

% ----------------------------------------------

\subsection{Token Copy Overhead Elimination}
\label{sec:moe:copy}

The SOTA MoE inference implementation directly uses grouped GEMM APIs, which usually require a contiguous layout of each input tensor.
However, for each expert in MoE, one of the input tensor, the token tensor, is a subset of the input token sequence.
The tokens in this subset are generally not consecutive, so the token tensor for GEMM input must be constructed specifically by gathering the tokens from the token sequence.
Every expert requires such a gether operation, and the copy overhead can be significant.

To eliminate this overhead, we introduce a token index array for every expert, containing the indices of the tokens routed to the expert.
Then the kernel only needs to load the token vectors from the original token sequence with the target token indices, rather than from the specifically prepared contiguous token tensors.
Atomic operations are used to scatter tokens into buckets corresponding to experts, which is the common technique in radix-based algorithms~\cite{li_topk_2024}.

% ----------------------------------------------

\subsection{Common GEMM Optimizations for Latest GPUs}
\label{sec:moe:gemm}

The most important metric for evaluating the performance of GEMM-like kernels is the computing power achieved on GPUs.
The latest GPUs offer increasingly higher hardware computing power with limited improvements in memory bandwidth.
As a result, to achieve the peak computing power of high-throughput Tensor Cores, several types of optimization must be comprehensively applied in our MoE kernel.
Moreover, we need to make full exploit of the new hardware features.

Specifically, we use the following GEMM optimizations.
\begin{itemize}
    \item We leverage asynchronous warpgroup level matrix multiply-accumulate (WGMMA) instructions to make full use of the Tensor Core computing power~\cite{nv_ptx}.
    \item We use asynchronous copy instructions to overlap the latency of memory accesses~\cite{nv_ptx}.
    \item We implement a two-stage pipeline for data prefetch and circular copy between global memory and shared memory to ensure the full load of Tensor Cores.
    \item We improve the L2 cache hit rate with the tile swizzle technique for GEMMs with large tiles.
\end{itemize}

%% file: tex/evaluation.tex
\section{Evaluation}
\label{sec:eval}

As mentioned in \Cref{sec:moe:gemm}, the main metric to evaluate the performance of our MoE kernel is the computing power achieved on GPUs.
We use two latest NVIDIA Hopper GPUs to evaluate our kernel:
\begin{itemize}
    \item H20 with peak FP16/BF16 Tensor Core throughput 146 TFLOPS;
    \item H800 with peak FP16/BF16 Tensor Core throughput 989 TFLOPS.
\end{itemize}

We use a token sequence length equal to 4096.
The shape of the expert weight tensor is [3584, 2560].
There are 64 experts in total, among which each token is routed to 8 experts.
We design three types of scenarios:
\begin{enumerate}
    \item balanced case: tokens are averagely routed to all experts;
    \item best case: all tokens are routed to the same 8 experts, i.e., we only need to compute 8 GEMMs;
    \item worst case: nearly all tokens are routed to the same 8 experts, but the other 56 experts each receive only one token, degrading these 56 GEMMs into extremely memory-bound cases.
\end{enumerate}

The results are shown in \Cref{tab:res}.

\begin{table}[ht]
    \centering
    \begin{threeparttable}
    \begin{tabular}{cccccc}
        \toprule
        \multirow{2}{*}{Case} & \multicolumn{2}{c}{\textbf{H20}} && \multicolumn{2}{c}{\textbf{H800}} \\
        \cmidrule{2-3}
        \cmidrule{5-6}
        & TFLOPS & peak\% && TFLOPS & peak\% \\
        \midrule
        Balanced & 138.23 & 94.67 && 838.87 & 84.82 \\
        Best & 138.55 & 94.89 && 897.03\tnote{1} & 90.70\tnote{1} \\
        Worst & 131.57 & 90.11 && 587.20 & 59.37 \\
        \bottomrule
    \end{tabular}
    \begin{tablenotes}
        \footnotesize
        \item[1] The best case on H800 uses a much larger sequence length and weight shape than the other settings, because the default sequence length and expert weight shape are too small to reach the peak throughput of H800.
    \end{tablenotes}
    \end{threeparttable}
    \caption{MoE inference kernel evaluation on NVIDIA H20 and H800}
    \label{tab:res}
\end{table}

From the results, we see that our MoE kernel achieves up to 95\% peak Tensor Core throughput in the balanced case on H20, and 85\% on H800.
In the best case, our kernel achieves up to 91\% of the peak throughput on H800.
Even in the worst case, our kernel achieves about 90\% peak throughput on H20, though its performance on H800 degrades to 60\% peak throughput on H800.
The best case and the worst case are both rare in practice, and the results in the balanced case provide a reference performance of our MoE kernel.

%% file: tex/conclusion.tex
\section{Conclusion}
\label{sec:conclusion}

We propose a static batching framework for general irregular workloads on GPUs with a novel task mapping algorithm.
We extend this framework to handle potential empty tasks in a batch and apply it to the MoE model inference.
In addition, we leverage several optimizations to implement a highly efficient MoE kernel, including expert ordering, token copy overhead elimination, and GEMM optimizations targeting the latest GPUs.
Evaluating on NVIDIA Hopper GPUs, our kernel achieves up to 95\% of the peak Tensor Core throughput on H20 and up to 91\% on H800.

%% file: main.bbl
\begin{thebibliography}{19}
\providecommand{\natexlab}[1]{#1}
\providecommand{\url}[1]{\texttt{#1}}
\expandafter\ifx\csname urlstyle\endcsname\relax
  \providecommand{\doi}[1]{doi: #1}\else
  \providecommand{\doi}{doi: \begingroup \urlstyle{rm}\Url}\fi

\bibitem[Chen et~al.(2023)Chen, Brock, Porumbescu, Buluc, Yelick, and Owens]{chen_atos_2023}
Yuxin Chen, Benjamin Brock, Serban Porumbescu, Aydin Buluc, Katherine Yelick, and John Owens.
\newblock {Atos}: A task-parallel {GPU} scheduler for graph analytics.
\newblock In \emph{Proceedings of the 51st International Conference on Parallel Processing}, ICPP '22, New York, NY, USA, 2023. Association for Computing Machinery.
\newblock ISBN 9781450397339.
\newblock \doi{10.1145/3545008.3545056}.
\newblock URL \url{https://doi.org/10.1145/3545008.3545056}.

\bibitem[DeepSeek-AI(2024)]{deepseek_v3_2024}
DeepSeek-AI.
\newblock Deepseek-v3 technical report, 2024.
\newblock URL \url{https://arxiv.org/abs/2412.19437}.

\bibitem[Duan et~al.(2024)Duan, Lu, Duanmu, Li, Zhang, Lin, Stoica, and Zhang]{duan_muxserve_2024}
Jiangfei Duan, Runyu Lu, Haojie Duanmu, Xiuhong Li, Xingcheng Zhang, Dahua Lin, Ion Stoica, and Hao Zhang.
\newblock {M}ux{S}erve: Flexible spatial-temporal multiplexing for multiple {LLM} serving.
\newblock In Ruslan Salakhutdinov, Zico Kolter, Katherine Heller, Adrian Weller, Nuria Oliver, Jonathan Scarlett, and Felix Berkenkamp, editors, \emph{Proceedings of the 41st International Conference on Machine Learning}, volume 235 of \emph{Proceedings of Machine Learning Research}, pages 11905--11917. PMLR, 21--27 Jul 2024.
\newblock URL \url{https://proceedings.mlr.press/v235/duan24a.html}.

\bibitem[Fedus et~al.(2022)Fedus, Dean, and Zoph]{fedus_reviewsparseexpertmodels_2022}
William Fedus, Jeff Dean, and Barret Zoph.
\newblock A review of sparse expert models in deep learning, 2022.
\newblock URL \url{https://arxiv.org/abs/2209.01667}.

\bibitem[Hejazi(2024)]{nv_groupedgemm_2024}
Babak Hejazi.
\newblock Introducing grouped {GEMM} {APIs} in {cuBLAS} and more performance updates, 6 2024.
\newblock URL \url{https://developer.nvidia.com/blog/introducing-grouped-gemm-apis-in-cublas-and-more-performance-updates/}.

\bibitem[Jiang et~al.(2024)Jiang, Sablayrolles, Roux, Mensch, Savary, Bamford, Chaplot, de~las Casas, Hanna, Bressand, Lengyel, Bour, Lample, Lavaud, Saulnier, Lachaux, Stock, Subramanian, Yang, Antoniak, Scao, Gervet, Lavril, Wang, Lacroix, and Sayed]{jiang_mixtralexperts_2024}
Albert~Q. Jiang, Alexandre Sablayrolles, Antoine Roux, Arthur Mensch, Blanche Savary, Chris Bamford, Devendra~Singh Chaplot, Diego de~las Casas, Emma~Bou Hanna, Florian Bressand, Gianna Lengyel, Guillaume Bour, Guillaume Lample, Lélio~Renard Lavaud, Lucile Saulnier, Marie-Anne Lachaux, Pierre Stock, Sandeep Subramanian, Sophia Yang, Szymon Antoniak, Teven~Le Scao, Théophile Gervet, Thibaut Lavril, Thomas Wang, Timothée Lacroix, and William~El Sayed.
\newblock Mixtral of experts, 2024.
\newblock URL \url{https://arxiv.org/abs/2401.04088}.

\bibitem[Kashi et~al.(2022)Kashi, Nayak, Kulkarni, Scheinberg, Lin, and Anzt]{knk_batch_2022}
Aditya Kashi, Pratik Nayak, Dhruva Kulkarni, Aaron Scheinberg, Paul Lin, and Hartwig Anzt.
\newblock Batched sparse iterative solvers on {GPU} for the collision operator for fusion plasma simulations.
\newblock In \emph{2022 {IEEE} International Parallel and Distributed Processing Symposium, {IPDPS} 2022}, pages 157--167, Lyon, France, May 30 -- June 3 2022. {IEEE}.
\newblock \doi{10.1109/IPDPS53621.2022.00024}.

\bibitem[Kosaian et~al.(2021)Kosaian, Phanishayee, Philipose, Dey, and Vinayak]{kpp_batch_2021}
Jack Kosaian, Amar Phanishayee, Matthai Philipose, Debadeepta Dey, and Rashmi Vinayak.
\newblock Boosting the throughput and accelerator utilization of specialized {CNN} inference beyond increasing batch size.
\newblock In Marina Meila and Tong Zhang, editors, \emph{Proceedings of the 38th International Conference on Machine Learning}, volume 139 of \emph{Proceedings of Machine Learning Research}, pages 5731--5741, Virtual Event, 18--24 Jul 2021. PMLR.
\newblock URL \url{https://proceedings.mlr.press/v139/kosaian21a.html}.

\bibitem[Kranen and Nguyen(2024)]{nv_moe_2024}
Kyle Kranen and Vinh Nguyen.
\newblock Applying {Mixture of Experts} in {LLM} architectures, 2024.
\newblock URL \url{https://developer.nvidia.com/blog/applying-mixture-of-experts-in-llm-architectures/}.

\bibitem[Li et~al.(2019)Li, Liang, Yan, Jia, and Li]{li_tiling_2019}
Xiuhong Li, Yun Liang, Shengen Yan, Liancheng Jia, and Yinghan Li.
\newblock A coordinated tiling and batching framework for efficient {GEMM} on {GPUs}.
\newblock In \emph{Proceedings of the 24th Symposium on Principles and Practice of Parallel Programming}, PPoPP '19, pages 229--241, New York, NY, USA, 2019. Association for Computing Machinery.
\newblock ISBN 9781450362252.
\newblock \doi{10.1145/3293883.3295734}.
\newblock URL \url{https://doi.org/10.1145/3293883.3295734}.

\bibitem[Li et~al.(2024)Li, Zhou, Zhang, Wei, Li, and Chen]{li_topk_2024}
Yifei Li, Bole Zhou, Jiejing Zhang, Xuechao Wei, Yinghan Li, and Yingda Chen.
\newblock {RadiK}: Scalable and optimized {GPU}-parallel radix top-k selection.
\newblock In \emph{Proceedings of the 38th ACM International Conference on Supercomputing}, ICS '24, pages 537--548, New York, NY, USA, 2024. Association for Computing Machinery.
\newblock ISBN 9798400706103.
\newblock \doi{10.1145/3650200.3656596}.
\newblock URL \url{https://doi.org/10.1145/3650200.3656596}.

\bibitem[Meta(2024)]{meta_llama31_2024}
Meta.
\newblock {Llama-3.1-405B-Instruct}, 2024.
\newblock URL \url{https://huggingface.co/meta-llama/Llama-3.1-405B-Instruct}.

\bibitem[NVIDIA(2024{\natexlab{a}})]{cuda_mps}
NVIDIA.
\newblock Multi-process service, 2024{\natexlab{a}}.
\newblock URL \url{https://docs.nvidia.com/deploy/mps/index.html}.

\bibitem[NVIDIA(2024{\natexlab{b}})]{nv_cublas_2024}
NVIDIA.
\newblock {cuBLAS}, 2024{\natexlab{b}}.
\newblock URL \url{https://docs.nvidia.com/cuda/cublas/index.html}.

\bibitem[NVIDIA(2025{\natexlab{a}})]{cuda_guide}
NVIDIA.
\newblock {CUDA} {C++} programming guide, 2025{\natexlab{a}}.
\newblock URL \url{https://docs.nvidia.com/cuda/cuda-c-programming-guide/index.html}.

\bibitem[NVIDIA(2025{\natexlab{b}})]{nv_ptx}
NVIDIA.
\newblock Parallel thread execution isa, 2025{\natexlab{b}}.
\newblock URL \url{https://docs.nvidia.com/cuda/parallel-thread-execution/index.html}.

\bibitem[Rajbhandari et~al.(2022)Rajbhandari, Li, Yao, Zhang, Aminabadi, Awan, Rasley, and He]{ms_dsmoe_2022}
Samyam Rajbhandari, Conglong Li, Zhewei Yao, Minjia Zhang, Reza~Yazdani Aminabadi, Ammar~Ahmad Awan, Jeff Rasley, and Yuxiong He.
\newblock {D}eep{S}peed-{M}o{E}: Advancing mixture-of-experts inference and training to power next-generation {AI} scale.
\newblock In Kamalika Chaudhuri, Stefanie Jegelka, Le~Song, Csaba Szepesvari, Gang Niu, and Sivan Sabato, editors, \emph{Proceedings of the 39th International Conference on Machine Learning}, volume 162 of \emph{Proceedings of Machine Learning Research}, pages 18332--18346. PMLR, 17--23 Jul 2022.
\newblock URL \url{https://proceedings.mlr.press/v162/rajbhandari22a.html}.

\bibitem[Tillet(2024)]{triton_groupedgemm_2024}
Philippe Tillet.
\newblock Tutorials: Group {GEMM}, 2024.
\newblock URL \url{https://triton-lang.org/main/getting-started/tutorials/08-grouped-gemm.html}.

\bibitem[Tzeng et~al.(2010)Tzeng, Patney, and Owens]{tzeng_irregular_2010}
Stanley Tzeng, Anjul Patney, and John~D. Owens.
\newblock Task management for irregular-parallel workloads on the {GPU}.
\newblock In \emph{Proceedings of the Conference on High Performance Graphics}, HPG '10, pages 29--37, Goslar, DEU, 2010. Eurographics Association.

\end{thebibliography}
